\documentclass[preprint2]{aastex63}

\usepackage{CJK}
\usepackage{amsmath}
\usepackage{bm}
\usepackage{multirow}
\usepackage{float}    
\usepackage{hyperref}
\usepackage{xcolor}

\newcommand{\petit}{\texttt{petitRADTRANS}}

\newcommand{\pmn}{\texttt{DYNESTY}}

\received{\today}
\revised{}
\accepted{}
\submitjournal{ApJ}

\shorttitle{HWO}
\shortauthors{Wang}
\graphicspath{{./}{figures/}}

\begin{document}
\begin{CJK*}{UTF8}{gbsn}

\title{On the Impact of Correlated Noise and Spectral Resolution on the Retrieval Analysis of the Habitable World Observatory}

\correspondingauthor{Ji Wang}
\email{wang.12220@osu.edu}

\author[0000-0002-4361-8885]{Ji Wang (王吉)}
\affiliation{Department of Astronomy, The Ohio State University, 100 W 18th Ave, Columbus, OH 43210 USA}

\author[0009-0000-7417-3201]{Philipp A. Huber}
\affiliation{ETH Zurich, Institute for Particle Physics and Astrophysics, Wolfgang-Pauli-Strasse 27, 8093 Zurich, Switzerland}
\affiliation{National Center of Competence in Research PlanetS (\url{www.nccr-planets.ch})}

\author[0000-0003-3829-7412]{Sascha P. Quanz}
\affiliation{ETH Zurich, Institute for Particle Physics and Astrophysics, Wolfgang-Pauli-Strasse 27, 8093 Zurich, Switzerland}
\affiliation{National Center of Competence in Research PlanetS (\url{www.nccr-planets.ch})}
\affiliation{ETH Zurich, Department of Earth Sciences, Sonneggstrasse 5, 8092 Zurich, Switzerland}

\begin{abstract}

Finding signs of life elsewhere in the universe is the holy grail of the field of exoplanets. Future space missions such as the Habitable World Observatory (HWO) are under development to search for biosignatures in exoplanets. We investigate the impact of correlated noise and spectral resolution on the retrieved biosignature chemical abundances. At the nominal spectral resolving power R=140 for HWO, we show that in 40\% of the simulated runs the retrieved biosignature (H$_2$O and O$_2$) abundances are at least 1-$\sigma$ off the input ground truth. At R=1000, the 1-$\sigma$ inaccuracy rate  drops to 10\%. We provide an empirical relationship between the retrieved biosignature abundance uncertainty and the amplitude of the correlated noise. As part of the mitigation plan to reduce the impact of correlated noise on retrieval accuracy at low spectral resolution, we investigate the synergy between the HWO and LIFE space missions that cover ultraviolet, optical, and thermal-infrared wavelengths. {{After considering clouds and their effect on planet albedo, we find that the two missions are complementary in that (1) more biosignatures (H$_2$O, CO$_2$, O$_2$, and O$_3$) are detectable with a broader wavelength coverage; (2) retrieval uncertainty improves with the joint HWO+LIFE data set; and (3) LIFE is more sensitive to the atmospheric temperature profile, surface pressure, and planet radius. This work provides evidence to support the choice of a medium resolution at R=1000 instead of R=140 for HWO and a quantitative relationship between the retrieved abundance uncertainty and the level of correlated noise at different spectral resolutions.      }}     

\end{abstract}



\section{Introduction}
\label{sec:intro}

Future space missions such as the Habitable World Observatory~\citep[HWO, ][]{Mennesson2024} and the LIFE space mission~\citep[LIFE, ][]{Quanz2022} are under development to detect habitable planets and search for biosignatures on those planets. Recent instrumental development has closed the gap between current technology and the ambitious scientific goal of detecting biosignatures~\citep{Mennesson2024}. Biosignatures of exoplanets are signs of life that can be remotely detected~\citep{Schwieterman2018}. For HWO particularly, previous studies have covered biosignatures such as gases in the atmosphere~\citep{Latouf2025}, polarimetric signal~\citep{Kenneth2024}, and red edge~\citep{Borges2024}.

However, biosignature detection faces several astrophysical challenges. For example, exo-zodiacal dust can produce sufficient scattered light to overwhelm the signal of a planet in the habitable zone~\citep{Coker2018,Currie2023}. While the presence of cloud would increase the planet's albedo and therefore increase the planet detection significance, clouds can also decrease the absorption depth of biosignature gases, making characterization of their abundances difficult~\citep{Wang2018JATIS,Kelkar2025}.

From the perspective of instrumentation, a detailed trade study is desirable in a complicated high-dimensional space including parameters such as wavelength coverage, spectral resolution, and detector noise.~\citet{Morgan2024} discusses the trade between ultraviolet (UV), optical, and near-infrared (NIR) wavelengths and conclude that optical wavelength is an optimal spot for starlight suppression and inner working angle (IWA).~\citet{Wang2017a} conducts a comprehensive trade study between spectral resolution and detector noise, and conclude that R$\sim$400-1000 is optimal for HWO-like space missions in the presence of detector noise. With the development of detector technology, several studies havem been conducted on zero-noise detectors and their impact on HWO performance~\citep{Howe2024,Steiger2024}. 

{{This work represents a true end-to-end simulation to address two critical questions in the trade study: (1) What is the impact of correlated noise on the accuracy of measuring biosignature abundances? and (2) How does this impact vary with spectral resolution? We consider uncorrelated white noise and correlated noise as described by a Gaussian process (GP). These noise components are considered and inferred in a Bayesian retrieval framework. In the contexts of high-contrast imaging and spectroscopy in space, this is an improvement from the cross-correlation approach used in~\citet{Wang2017a} and the retrieval studies that consider only white noise~\citep[e.g., ][]{Feng2018,Alei2024}. }}

{{~\citet{Huber2025} demonstrates spectrally correlated noise can be reduced in synthetic measurements of LIFE given knowledge of the noise covariance matrix. This can be potentially applied to HWO. However, the source of correlated noise is different: instrumental perturbations are the major contributor for LIFE and speckle spectral noise is the major contributor for HWO~\citep{Wang2017a}. It remains unclear if the LIFE approach would be effective for HWO because speckle variation is both time-dependent and detector-location-dependent. Here, we adopt a general GP framework to be more widely applicable. 

Alternatively, auto-regressive models are used in analyzing astronomical data with correlated noise. For example, ~\citet{Caceres2019,Melton2024} used the auto-regressive integrated moving average (ARIMA) method to search for transiting planets in TESS and Kepler data. ~\citet{Farr2018} used GP-based continuous-time autoregressive moving-average method to model asteroseismological and planetary signals.~\citet{Ejaz2026} used auto-regressive models to understand the impact of correlated noise on false positives in analyzing radial velocity data. While these auto-regressive models are primarily used for time-series data with correlated noise, they may also be effective in analyzing spectral data, in addition to the GP approach used in this work.   

}}

We organize the paper as follows. \S \ref{sec:method} describes our methodology. \S \ref{sec:major_findings} presents major findings for HWO on the impact of correlated noise and spectral resolution on the accuracy of retrieved biosignature abundances. \S \ref{sec:hwo_life} discusses the synergy between HWO and LIFE. A summary of the paper can be found in \S \ref{sec:summary}.

\section{Method}
\label{sec:method}
\subsection{Generating an Earth-like Spectrum}
\label{sec:simulation}

We generate reflected-light spectra of an Earth-like planet using \petit~\citep{Molliere2019}. {{The parametrization is similar to~\citet{Feng2018} whose albedo spectrum model is based on~\citet{Cahoy2010} and noise model is based on~\citet{Robinson2016}. We use the same input values for the following overlapping parameters: surface pressure, surface gravity, planet radius, and volume mixing ratios for H$_2$O and O$_2$. 

We choose an albedo value of 0.3, which is similar to Earth's Bond albedo and different from the albedo value used in~\citet{Feng2018}. The Bond albedo value we choose is more relevant to the investigation in \S \ref{sec:hwo_life} on the energy budget of an Earth-like planet. We do not include clouds in the HWO simulation and investigation in \S \ref{sec:major_findings} because we would like to separate the retrieval of biosignature abundances and the retrieval of cloud  parameters. However, a cloudy case is considered when studying the synergy between UV+optical (HWO) and thermal infrared (LIFE) observations (see \S \ref{sec:hwo_life}). In that section, our cloud parameterization is different from~\citet{Feng2018} because their 4-parameter cloud parameterization (cloud-top pressure, thickness, optical depth, and coverage fraction) is not a default parameterization in\ \petit. All parameters used are listed in Table \ref{tab:retrieval_params}. }}

{{Furthermore, we note two more differences}}. First, the O$_3$ opacity database in \petit\ does not include ultraviolet data, so we omit O$_3$ in our HWO simulation. Second, we consider a wavelength coverage between 0.6 and 1.0 $\mu$m to focus on the study of biosignature abundance retrieval for H$_2$O and O$_2$ for HWO, although we consider a wavelength coverage between 0.3 and 1.0 $\mu$m for the HWO-LIFE synergy study  (more details in \S \ref{sec:hwo_life}). 

\begin{figure*}[th!]
\begin{tabular}{c}
\includegraphics[width=16.0cm]{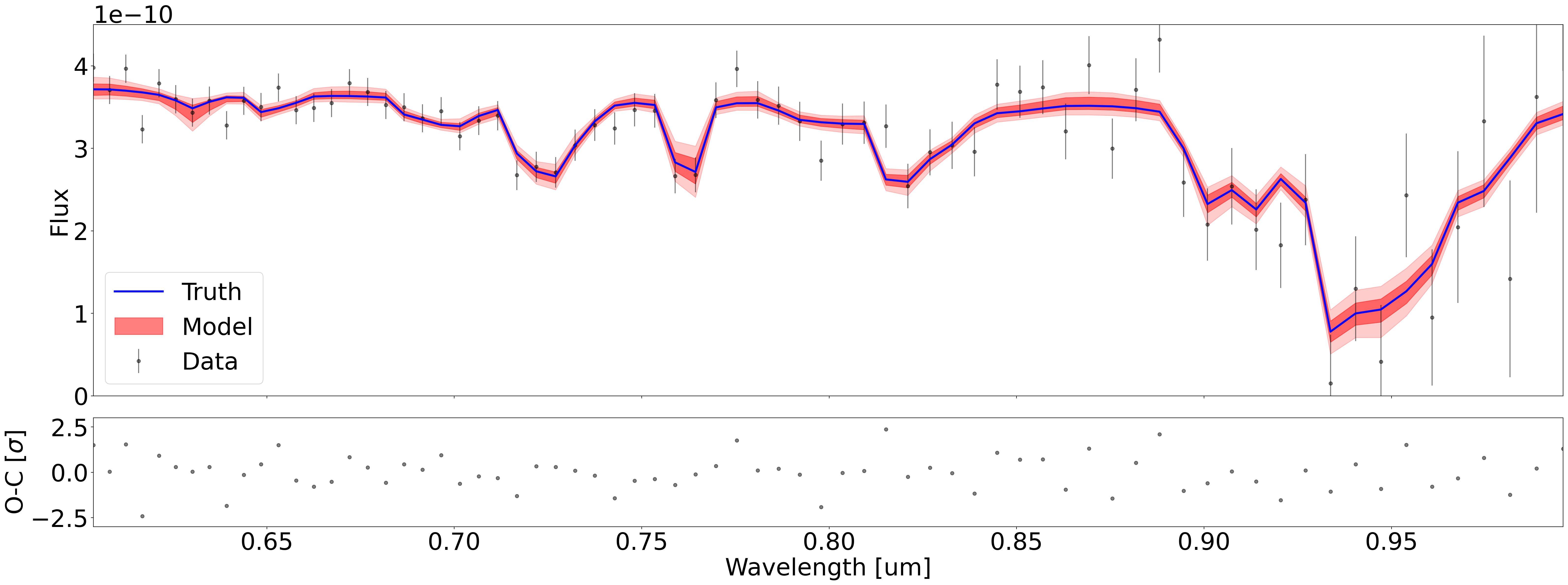}
\end{tabular}
\caption{{\bf{Simulated spectral data and retrieved Earth spectrum. }} Top: mock data with R = 140 without correlated noise are shown in black. These data are generated based on methods described in \S \ref{sec:method}. The blue solid line is the input noiseless spectrum. The red filled lines are 1-$\sigma$ (darker red) and 2-$\sigma$ (lighter red) distribution of the retrieved spectra from 100 posterior samples. Bottom: Residual between the data and model as measured by 1-$\sigma$ uncertainty. Results for R=1000 without correlated noise are qualitatively similar and therefore are not included in figures.    }
    \label{fig:model_data_feng2018}
\end{figure*} 

\begin{figure*}[th!]
    \centering
    \includegraphics[width=0.95\linewidth]{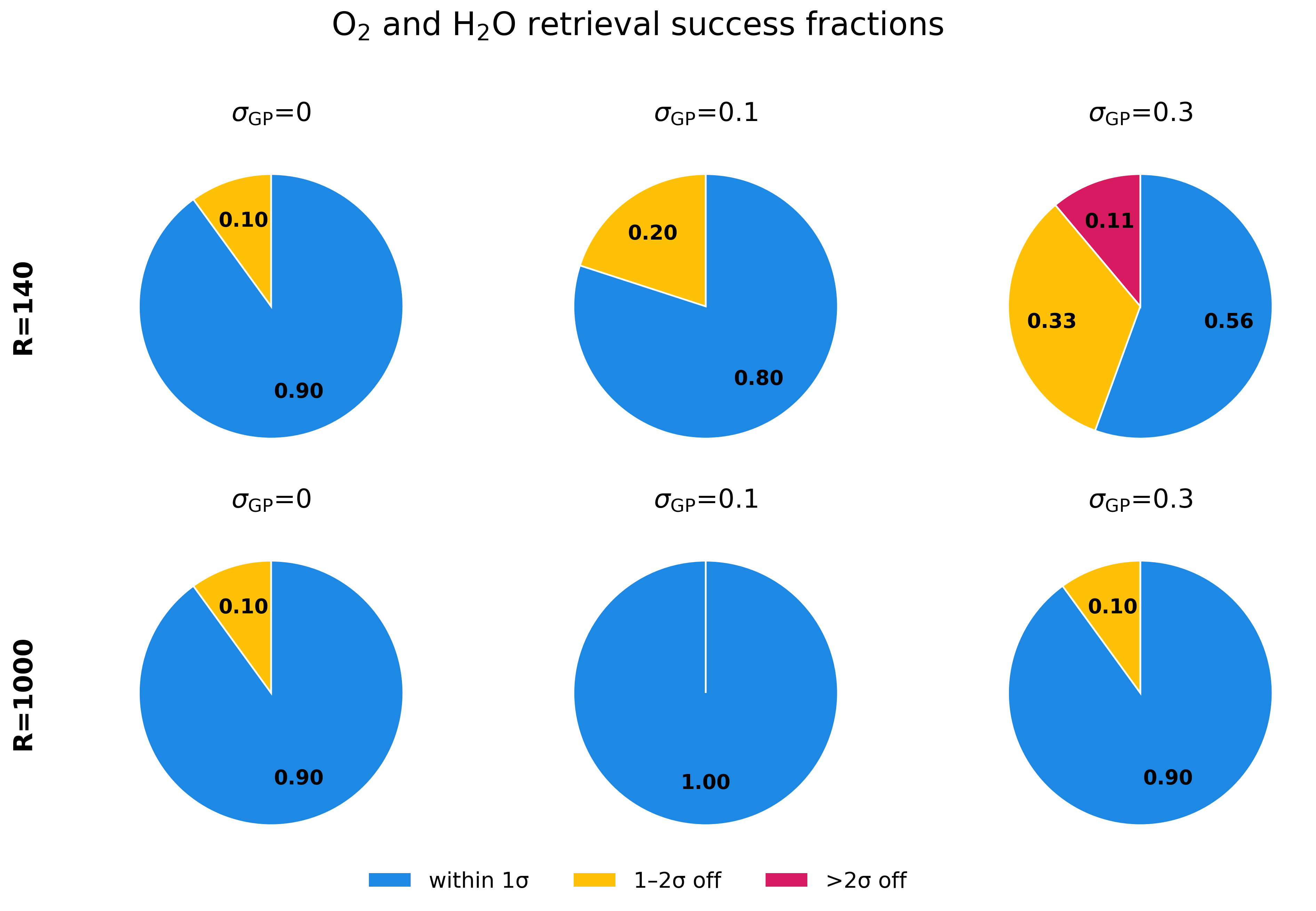}
    \caption{{\bf Higher spectral resolution ($R$=140 to 1000) significantly improves biosignature retrieval success rates, particularly in the presence of correlated noise.} Fractions of successful $\mathrm{O_2}$ and $\mathrm{H_2O}$ retrievals across different spectral resolutions (rows) and Gaussian Process (GP) noise amplitudes (columns). Retrievals are categorized by whether the posterior medians fall within $1\sigma$ (blue), $1\text{--}2\sigma$ (yellow), or $>2\sigma$ (red) of the true input values.}
    \label{fig:o2_h2o_success_pie}
\end{figure*}


{{We use \petit\ to generate scattering and emission spectra for an Earth-like planet. We then divide the generated spectra by a stellar spectrum to remove the stellar component to generate a reflected-light spectrum.}} This allows us to compare our reflected-light spectrum to those in~\citet{Feng2018} obtained from Earth-orbiting satellites. {{We choose a scaling factor to match the planet-star flux ratio so that the reflection is at $\sim3.5\times10^{-10}$ at spectral continuum (Fig. \ref{fig:model_data_feng2018}). This is comparable to the continuum level of the spectrum as shown in Fig. 5 in~\citet{Feng2018} between 0.6 and 1.0 $\mu$m.}} 

\subsection{Simulating Noise}
\label{sec:simulation_noise}

{{We add uncorrelated Gaussian noise (i.e., white noise) to the synthetic spectrum.}} Similar to~\citet{Feng2018}, we use a wavelength dependent signal-to-noise ratio (SNR). At R=140, we set SNR = 20 for wavelength between 0.5 and 0.7 $\mu$m. SNR drops linearly to 50\% from the maximum SNR at 0.3 $\mu$m and to 10\% of the maximum SNR to the red end at 1.0 $\mu$m.

Low-resolution data are likely compromised by speckle spectral noise~\citep{Wang2017a}. Even for a flat spectral energy distribution (SED) of an input spectrum, the wavefront control system would distort the output SED such that it mimics broad absorption bands of biosignatures~\citep[see Fig. 14 in ][]{Wang2017a}. This is because the wavefront control system is optimized at a certain wavelength. Away from the optimal wavelength, starlight suppression becomes worse. As a result, the starlight leakage causes an increase of flux away from the optimal wavelength, resulting in an absorption-like feature. 

{{To quantify the impact of speckle spectral noise at $R=140$, we compare retrievals of data with uncorrelated white noise and data with correlated noise}}. The correlated noise is approximated by a GP with a squared exponential kernel: 
\begin{equation}
K_{ij}=
\sigma_{{\rm w},i}^{2}\delta_{ij}
+
\sigma_{\rm GP}^{2}
\exp\left[
-\frac{(\lambda_i-\lambda_j)^2}
{2\lambda_{\rm GP}^{2}}
\right],
\label{eq:sq_ex}
\end{equation}
{{where $\sigma_{{\rm w},i}$ is the white-noise uncertainty in the $i$th wavelength channel, $\delta_{ij}$ is the Kronecker delta, $\sigma_{\rm GP}$ is the amplitude of the correlated component, and $\lambda_{\rm GP}$ is its correlation length. The white-noise contribution therefore appears only in the diagonal elements of the covariance matrix. We also infer GP parameters with the same parametrization in the retrieval analysis.}} The default GP parameters are listed in Table~\ref{tab:retrieval_params}. The adopted GP correlation length and amplitude are consistent with the correlated-noise properties found by \citet{Wang2017a}. In terms of chemical species, we focus on H$_2$O and O$_2$, the two biosignature gases that are measurable in our HWO analysis.

\subsection{Setting Up Retrieval Analyses}
\label{sec:simulation_retrieval}

We use \pmn~\citep{Speagle2020} to perform the retrieval analysis in a Bayesian framework. The free parameters and their true values are given in Table \ref{tab:retrieval_params}, as well as the priors. We use 2000 live points and a stopping criterion that the change in log evidence is less than 0.1 in successive iterations or the total number of effective posterior samples has reached 100,000.

\begin{figure*}[th!]
\begin{tabular}{c}
\includegraphics[width=16.0cm]{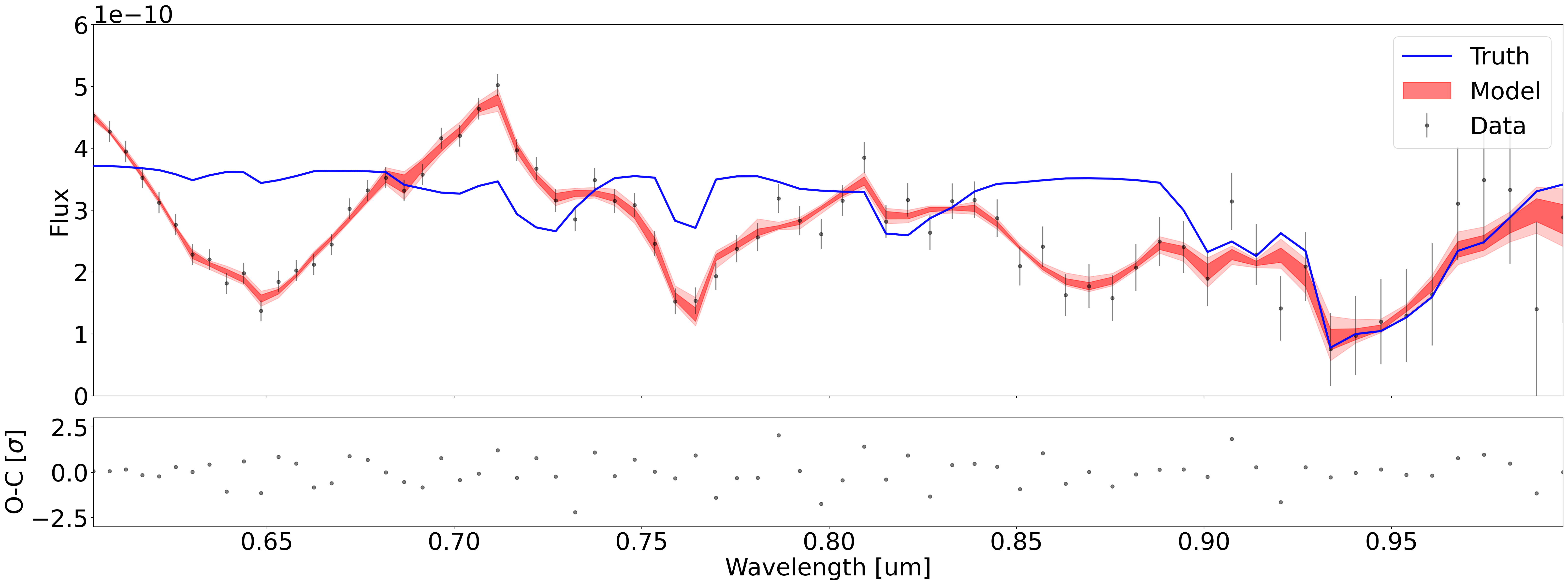}
\end{tabular}
\caption{{\bf{Simulated spectral data and retrieved Earth spectrum in the presence of correlated noise. }} Top: mock data with R = 140 with spectral correlated noise are shown in black. The blue solid line is the input noiseless spectrum. The red filled lines are 1-$\sigma$ (darker red) and 2-$\sigma$ (lighter red) distribution of the retrieved spectra from 100 posterior samples. The mismatch between the red and blue curves is due to the added correlated noise that cause low-frequency variation, which could confuse biosignature detection and inference at lower spectral resolution (see Fig. \ref{fig:corner_r140_gp}). Bottom: Residuals between the data and the model as measured by 1-$\sigma$ uncertainty. Results for higher spectral resolution at R=1000 are discussed in \S \ref{sec:r1000}.    }
    \label{fig:model_data_r140_gp}
\end{figure*} 

\begin{figure*}[bh!]
\begin{tabular}{c}
\includegraphics[width=16.0cm]{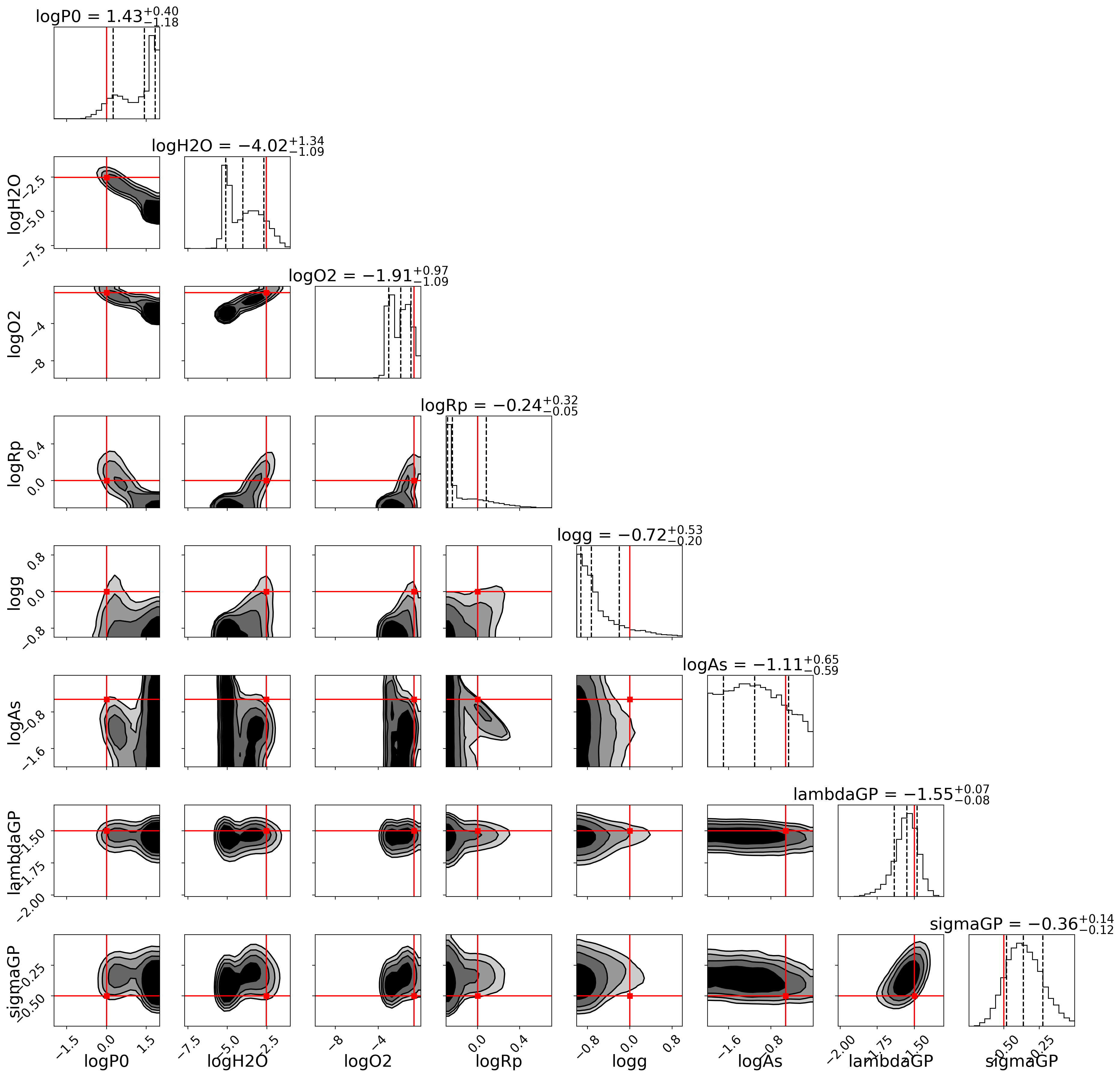}
\end{tabular}
\caption{{{Corner plot for the retrieval for mock data with R=140 in the presence of correlated noise.}} Contours from inside out are for 0.5, 1.0, 1.5, and 2.0-$\sigma$ levels. Red lines are input values for generating the mock data.  }
    \label{fig:corner_r140_gp}
\end{figure*} 

\section{Major Findings for HWO}
\label{sec:major_findings}

\subsection{Correlated Noise Decreases Retrieval Accuracy of H$_2$O and O$_2$}
\label{sec:correlated_noise}

At $R=140$, spectral correlated noise both increases the failure rate and broadens the inferred uncertainty for H$_2$O and O$_2$. When considering H$_2$O and O$_2$ together, the 1-$\sigma$ failure rate increases from 10\% to 40\% for inferring biosignature abundances in the presence of correlated noise. The fractions of successful and unsuccessful retrievals are summarized in Fig. \ref{fig:o2_h2o_success_pie} for different correlated noise amplitudes at various spectral resolutions. Fig. \ref{fig:model_data_r140_gp} gives an example of the impact of added correlated noise on the appearance of the simulated spectrum. Fig. \ref{fig:corner_r140_gp} shows that H$_2$O and O$_2$ abundances can be 1-2 $\sigma$ off the input values in the presence of correlated noise. 

For H$_2$O individually, among the 10 simulations, 90\% of the retrieved H$_2$O abundances are consistent with the input value within 1-$\sigma$ and none are off by more than 2-$\sigma$ (Table~\ref{tab:r140_snr20_nogp}). The posterior uncertainty typically has a characteristic width of roughly 0.4--0.8 dex. Once correlated noise is included, the retrieval accuracy decreases. Only 67\% of the retrieved H$_2$O abundances remain consistent with the input value within 1-$\sigma$ and 10\% are off by more than 2-$\sigma$ (Table~\ref{tab:r140_snr20}). In addition, the posterior uncertainty increases substantially, typically to about 0.8--1.3 dex. 

A similar trend is seen for O$_2$. Without correlated noise, 90\% of the retrieved O$_2$ abundances are consistent with the input value within 1-$\sigma$ (Table~\ref{tab:r140_snr20_nogp}). The corresponding posterior uncertainty is typically about 0.3--0.7 dex. With correlated noise included, the fraction of successful O$_2$ retrievals within 1-$\sigma$ decreases to 67\% (Table~\ref{tab:r140_snr20}). Meanwhile, the typical uncertainty increases to roughly 0.7--1.2 dex.

\subsection{Higher Spectral Resolution Reduces the Impact of Correlated Noise}
\label{sec:r1000}

We next investigate how increasing the spectral resolution from $R=140$ to $R=1000$ affects the retrieval results. We choose R=1000 for the following two reasons. First, ~\citet{Wang2017a} concluded that the optimal spectral resolution is between R=400 and R=1000 for a 4-m to 12-m space mission in the presence of detector noise. Second, ~\citep{Ruffio2026} concluded that R$>$1000 is likely required to suppress the impact of correlated noise. Therefore, R=1000 represents a middle ground between the two previous studies. 

Similar to the $R=140$ case, we consider mock observations both with and without correlated speckle spectral noise. At $R=1000$, spectral lines are better resolved and the retrieval is less susceptible to confusion between true absorption features and instrumental systematics. Therefore, the main purpose of this section is to evaluate whether increasing spectral resolution can both reduce retrieval uncertainties and alleviate the negative impact of correlated noise on the inferred abundances of H$_2$O and O$_2$.

For the correlated-noise case, we follow the same treatment as described in \S \ref{sec:correlated_noise}, except that the spectral resolution is increased from 140 to 1000. The SNR per spectral channel at R=1000 is estimated based on the framework in~\citet{Wang2017a}. We first tune our end-to-end simulation code to produce an SNR of 20 for R=140 to be consistent with the SNR in \S \ref{sec:correlated_noise}. Then we change R from 140 to 1000 while maintaining the same exposure time in the simulation code to estimate the SNR. At R=1000, SNR per spectral channel is 7.5. At this higher spectral resolution, photon-noise still dominates and detector noise and other noise terms are not the major contributor to the noise budget.

We first compare the correlated-noise and white-noise cases at the same spectral resolution of $R=1000$. In the absence of correlated noise, 90\% of the retrieved H$_2$O abundances are consistent with the input value within 1-$\sigma$ with no cases beyond 2-$\sigma$ (Table \ref{tab:r1000_snr7p5_nogp}). When correlated noise is included, the statistics remain the same: 90\% of the retrieved H$_2$O abundances are consistent with the input value within 1-$\sigma$ and none are off by more than 2-$\sigma$ (Table \ref{tab:r1000_snr7p5_gp}). Therefore, the success rate for H$_2$O retrieval is unchanged by the presence of correlated noise at $R=1000$. A similar trend is seen for O$_2$ (Table \ref{tab:r1000_snr7p5_nogp}). The dominant effect of correlated noise is to broaden the posterior uncertainty. 

Compared to the $R=140$ results, this demonstrates that correlated noise affects the abundance inference much less severely at higher spectral resolution (see also Fig. \ref{fig:o2_h2o_success_pie} for a summary).

In the presence of correlated noise, increasing the spectral resolution substantially improves both the accuracy and precision of the H$_2$O and O$_2$ retrievals. We compare the correlated-noise cases between $R=140$ and $R=1000$. The typical uncertainty shrinks from roughly 0.8--1.3 dex to about 0.5--0.8 dex. For O$_2$, the typical uncertainty also decreases, from about 0.7--1.2 dex to roughly 0.4--0.8 dex. The same overall trend is seen when comparing the white-noise cases between $R=140$ and $R=1000$. 

Overall, the comparison shows that higher spectral resolution reduces retrieval uncertainties and alleviates the negative impact of correlated speckle spectral noise. This implies that the resolved molecular features at $R=1000$ provide sufficient information to distinguish true atmospheric absorption from instrumental systematics more robustly, making correlated noise substantially less damaging than in the low-resolution case. This finding is also consistent with another study that uses a more analytical approach~\citep{Ruffio2026}.  

\subsection{The Impact of Correlated-Noise Amplitude on Biosignature-Abundance Uncertainty}
\label{sec:amplitude}

\begin{figure*}[t]
    \centering
    \includegraphics[width=0.95\linewidth]{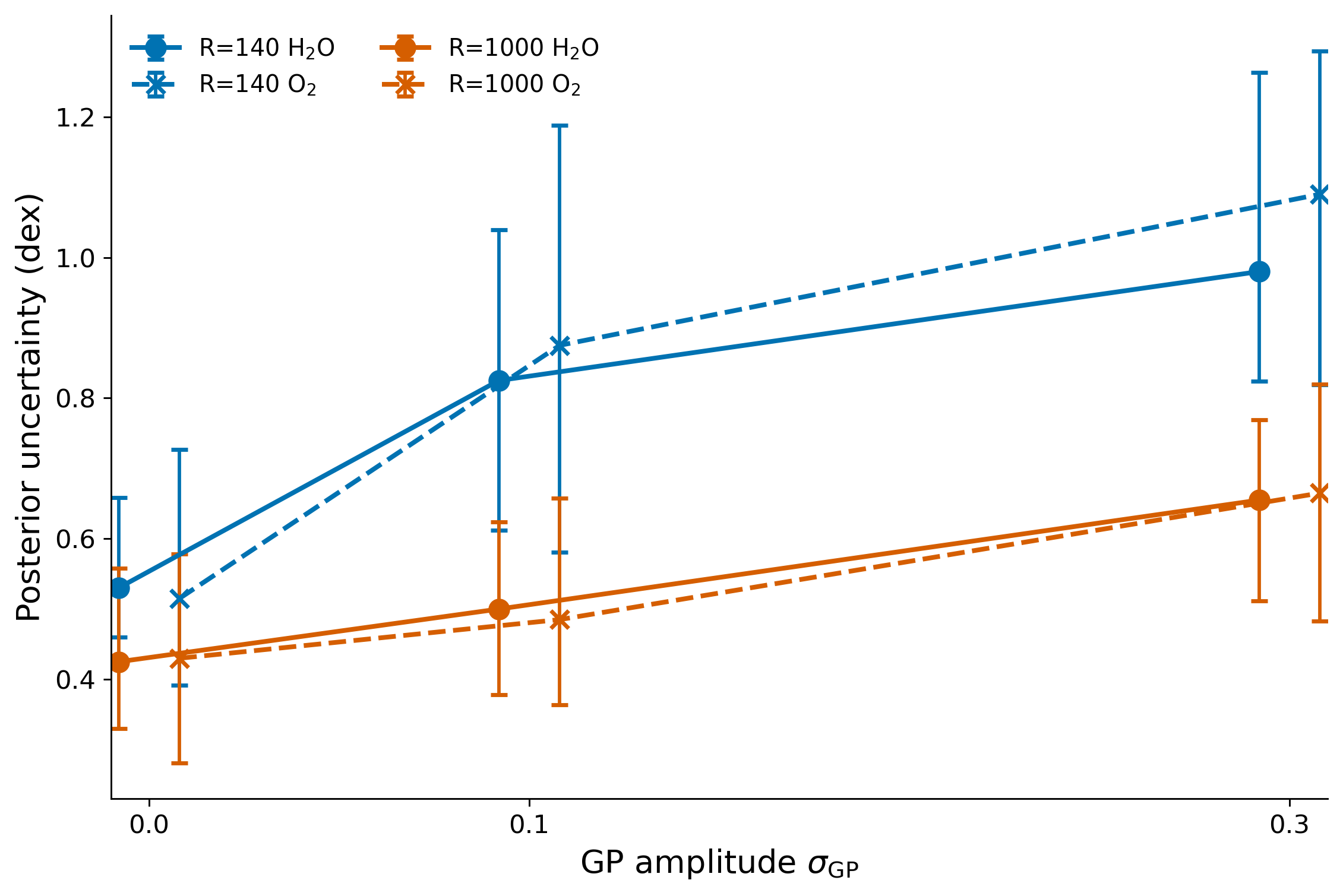}
    \caption{{\bf{Higher spectral resolution ($R$=140 to 1000) reduces the inferred biosignature uncertainties and the impact of correlated noise.}} Median posterior uncertainties of H$_2$O and O$_2$ as a function of the amplitude of correlated Gaussian noise. Blue symbols show the $R=140$ cases and red symbols show the $R=1000$ cases. Circles denote H$_2$O and crosses denote O$_2$. A small horizontal offset is applied between the H$_2$O and O$_2$ points for clarity. }
    \label{fig:gp_amp_vs_error}
\end{figure*}

To understand the impact of the amplitude of the correlated noise on the retrieval uncertainty, we conduct retrieval analyses by adding one more case in which the GP amplitude is 0.1. This is in comparison to the previous cases in which the GP amplitude is at 0.0 (the uncorrelated noise case) and 0.3. The simulation and retrieval setups are identical to previous cases. 

Fig. \ref{fig:gp_amp_vs_error} shows a comparison among different cases and for different species.  For each species and each noise amplitude, the plotted value is the median of all upper and lower posterior error bars from the ensemble of retrieval runs, and the vertical error bars indicate the 16th--84th percentile range of those pooled uncertainties. The GP amplitudes of 0.0, 0.1, and 0.3 correspond to the white-noise case and two correlated-noise cases, respectively. 

Posterior uncertainties increase with correlated-noise amplitude. This is consistent with physical intuition. However, at $R=140$, the uncertainties are more sensitive to the correlated noise as indicated by the sudden jump between 0.0 and 0.1 in amplitude. In comparison, at $R=1000$, the impact of correlated noise remains log-linear between 0.1 and 0.3 in GP amplitude. This demonstrates that higher spectral resolution reduces the impact of correlated noise on the retrieval of the biosignature gases such as H$_2$O and O$_2$.

\begin{figure*}[t]
    \centering
    \includegraphics[width=\linewidth]{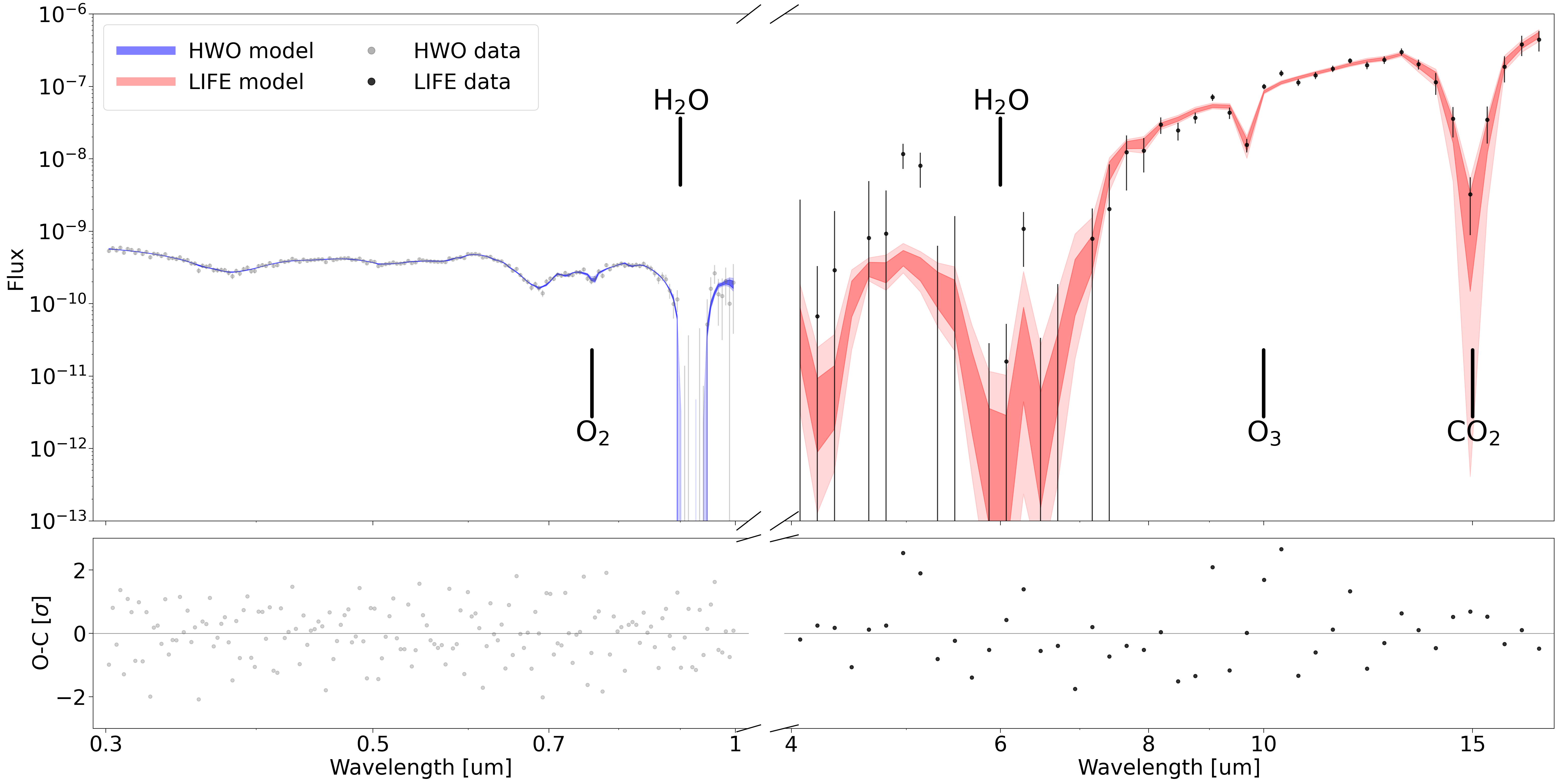}
    \caption{Top: Best-fit model spectra and residuals for the HWO and LIFE datasets (grey and black data points with error bars). Shaded regions show posterior predictive intervals, and the lower panel shows residuals in units of $\sigma$. Bottom: Residual between the data and model as measured by 1-$\sigma$ uncertainty. }
    \label{fig:model_data_hwo_life}
\end{figure*}

\section{The HWO-LIFE Synergy}
\label{sec:hwo_life}

Broadening the wavelength coverage can mitigate the inaccuracy in retrieving biosignature abundances. Moreover, extending the wavelength coverage from UV and optical wavelengths to the thermal-infrared wavelength can naturally increase the inventory of detectable biosignatures and constrain atmospheric properties such as temperature profile and clouds. This section places the HWO results for R=140 in a broader multi-wavelength retrieval context when combined with complementary thermal-infrared observations from the LIFE space mission. 

\subsection{Differences in Parametrization}

There are several differences in the parameterization adopted for the HWO--LIFE synergy study relative to the HWO-only study in \S~\ref{sec:major_findings}.

First, we include CO$_2$ and O$_3$ in the gas mixture because both produce prominent features in the thermal infrared. We note, however, that no UV opacity for O$_3$ is currently available in the \petit\ opacity tables. As a result, O$_3$ is not expected to be constrained in the HWO-only retrievals even though O$_3$ should be observed in actual HWO data. Although other chemical species could be included in the study, it is sufficient to include CO$_2$, O$_3$, and H$_2$O for the purpose of investigating and demonstrating retrieval precision and accuracy for LIFE.

Second, we include three additional cloud parameters: the cloud fraction ($f_c$) and two parameters associated with the power-law cloud description implemented in \petit. The power-law opacity corresponds to the cloud opacity at 0.35~$\mu$m and is sampled in log space. The power-law coefficient describes the wavelength dependence, including the trend and slope of the cloud opacity. Because cloud fraction is correlated with albedo, we assume a linear relation between the two, anchored such that the albedo is 0.1 when $f_c = 0$ and 0.9 when $f_c = 1$. 

Third, we add correlated noise only to the HWO data because it is the primary focus of this work. In addition, a conceptual framework has been proposed and demonstrated to calibrate and remove the correlated noise for the LIFE space mission~\citep{Huber2025}. In principle, a GP treatment of correlated noise could also be applied to the simulated LIFE data; however, we confirm that the retrieved amplitude of correlated noise is consistent with zero for the ``whitened'' LIFE data.

Lastly, we include a power-law index to describe the atmospheric temperature profile. The equilibrium temperature is calculated as a function of albedo:
\begin{equation}
T_{\rm eq} = T_{\oplus}
\left( \frac{1 - A}{1 - A_{\oplus}} \right)^{1/4},
\label{eq:teq_albedo}
\end{equation}
where $T_{\oplus} = 270$ K and $A_{\oplus} = 0.3$. The temperature profile is then computed using a power-law relation with a power-law index $\alpha$ (see Table \ref{tab:retrieval_params}), anchored at 1 bar with the equilibrium temperature. In reality, the temperature profile is more complicated than we assume and it can significantly impact the retrieved abundances~\citep[e.g.,][]{Konrad2024}. Here we assume a simplistic temperature profile that is non-isothermal in order to model spectral features in thermal emission. An example of the HWO and LIFE data and the retrieved spectra are given in Fig. \ref{fig:model_data_hwo_life}.

\subsection{Findings for the HWO-LIFE Synergy}
\label{sec:life_hwo_quant}

\begin{figure*}[t]
    \centering
    \includegraphics[width=\linewidth]{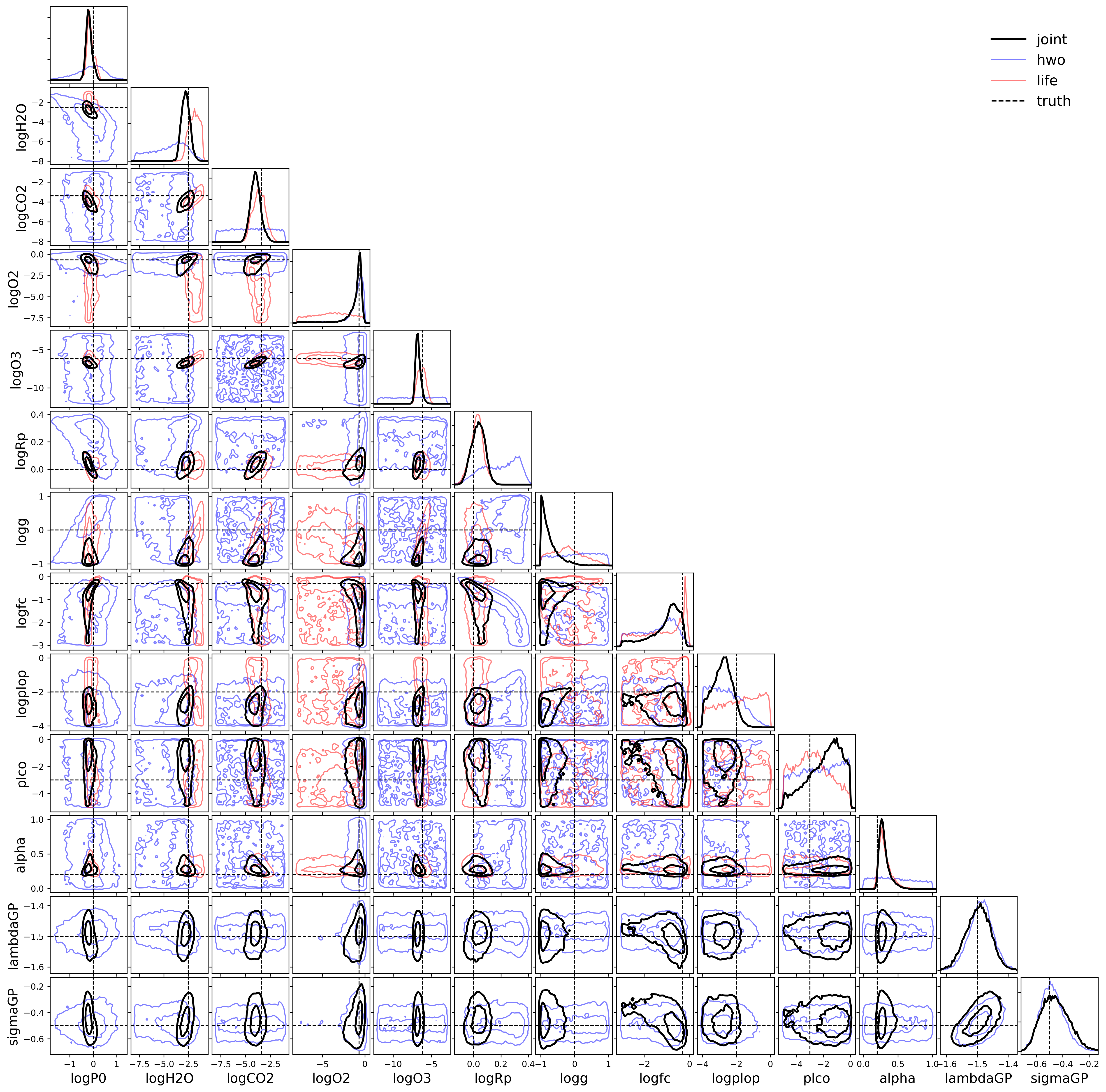}
    \caption{{\bf{HWO and LIFE are synergistic in measuring biosignature abundances and characterizing atmospheric and cloud properties.}} Corner plot comparing posterior constraints from the joint (black), HWO-only (blue), and LIFE-only (red) retrievals. Contours show the 1$\sigma$ and 2$\sigma$ credible regions. Black dashed lines mark the input truth values. }
    \label{fig:corner_overlay_joint_hwo_life}
\end{figure*}

Our findings are broadly consistent with those of~\citet{Alei2024}, further supporting the value of combining UV+optical reflected-light data with thermal-emission data for biosignature searches and for characterizing the atmospheres of Earth twins. {{Fig. \ref{fig:corner_overlay_joint_hwo_life} provides a visual comparison of posterior constraints and Table \ref{tab:hwo_life_joint_constraints} quantifies the retrieval results. We report our main results as follows. }}

{{In terms of the biosignature gases (H$_2$O, CO$_2$, O$_2$, and O$_3$), the retrieval using the joint HWO+LIFE data set returns smaller uncertainties than the retrievals using either individual data set alone. The most noticeable improvements are for H$_2$O and O$_3$, for which the uncertainties are approximately 1.5--2 times narrower. This is likely due to the simultaneous presence of spectral lines from the UV to the thermal IR and/or other better-constrained parameters (e.g., the cloud coverage factor) that lead to smaller uncertainties. Interestingly, the smaller posterior uncertainties also result in 1--2-$\sigma$ biased inferences for CO$_2$ and O$_3$.

Comparatively, HWO contributes the most to the O$_2$ detection. Assuming that the HWO+LIFE uncertainty results from independent contributions from HWO and LIFE, HWO contributes approximately 90\% of the O$_2$ constraint. {{We estimate the relative contribution using inverse-variance weighting, such that the contribution from a single mission (HWO or LIFE) is proportional to $1/\sigma^2$. Because the posterior uncertainties are asymmetric, we calculate the contribution separately using the upper and lower uncertainties and then take the mean of the two values.}} Similarly, LIFE contributes more strongly to the CO$_2$ and O$_3$ constraints, accounting for approximately 93\% and 97\%, respectively. HWO should also be sensitive to O$_3$; however, we cannot practically quantify its contribution because \petit\ does not currently include the UV opacity of O$_3$.

In addition, LIFE yields tighter constraints on the surface pressure and the power-law index describing the atmospheric temperature profile, contributing approximately 93\% and 94\% of the respective constraints. This is expected because thermal-emission spectra probe absorption features and are more directly sensitive to the atmospheric temperature structure than the reflected-light spectra obtained with HWO. LIFE also provides a more precise and unbiased measurement of the planet radius.

The joint HWO+LIFE retrieval yields biased values at greater than $2\sigma$ for $\log g$ and $\alpha$. {{In the case of $\log g$, the HWO and LIFE data sets alone return flat posterior distributions, indicating a lack of constraining power on this parameter. However, the joint data set returns a posterior peak that is inconsistent with the true input value. The biased $\log g$ retrieval may be mitigated by the accurate measurement of the planet radius from the LIFE mission and a well-characterized mass--radius relationship for rocky planets~\citep{Adibekyan_2021,Brinkman2025}. This issue may also be addressed more directly through future extreme-precision radial-velocity measurements. For $\alpha$, the T--P profile power-law index, it is mainly constrained by the LIFE data set, but the slightly narrower posterior distribution pushes the median value to more than $2\sigma$ away from the input $\alpha$ value.
}} These cautionary results underscore the need for careful scrutiny when interpreting these parameters.  
 }}

\subsection{Prospects for Methane (CH$_4$) Detection}
\label{sec:ch4}

The simultaneous detection of O$_2$ and CH$_4$ can be a strong indication of life for a habitable planet with a modern Earth atmosphere~\citep{Stueken2020, Crouse2021}. 

For HWO, wavelength coverage and IWA can both pose challenges in detecting CH$_4$. If HWO can only reach 1.0 $\mu$m on the red end, then CH$_4$ features will not be in the spectral coverage. Even if the HWO coverage extends to 1.8 $\mu$m, only an upper limit on the CH$_4$ abundance can be obtained~\citep{Alei2024} in an optimal cloud-free condition. The challenge is also highlighted in~\citet{Wang2018JATIS} for various cloud conditions. Methane features begin to become richer and deeper past 2$\mu$m. Observing beyond 2$\mu$m is challenging for both the system cooling and IWA. The IWA is 68 mas at 2$\mu$m for a 6-m HWO, which will significantly limit the number of searchable habitable planets within 10 pc that have angular separation of about 100 mas. 

{{For the LIFE mission, the major challenge is the loss of sensitivity towards the shorter wavelengths, i.e., the increased error bars as shown in Fig. \ref{fig:model_data_hwo_life}. The SNR at $\sim$4 $\mu$m where methane has strong spectral features is around 1 at R$\sim$30. However, ~\citet{Konrad2024} points out that the 7.7-$\mu$m methane feature can be distiguished from the water feature and detected with a SNR of 20 at R=100. As noted in~\citet{Huber2025}, the choice of a spectral resolution lower than R=100 was made for computational reasons and not a limitation of the hardware on LIFE. The gap in SNR can be filled by increasing the exposure time to longer than 10 days and/or improving the total throughput of the instrument. The requirement for detecting CH$_4$ will be further investigated in future work to drive the design of HWO and LIFE. }}

\section{Summary}
\label{sec:summary}

We investigate the impact of correlated noise on the accuracy of spectral retrieval for biosignatures (H$_2$O and O$_2$) for HWO (\S \ref{sec:major_findings}). We find that at R=140, 40\% of the mock runs retrieved H$_2$O and/or O$_2$ abundances that are off by at least 1-$\sigma$ even though the correlated noise is accurately modeled by a Gaussian process. 

A higher spectral resolving power (R=1000) significantly alleviates the issue: we find that only 10\% of the mock runs retrieved H$_2$O and/or O$_2$ abundances that are off by at least 1-$\sigma$. The result is summarized in Fig. \ref{fig:o2_h2o_success_pie}. 

Moreover, we also find that a higher spectral resolution is more robust against the impact of correlated noise in terms of the measurement uncertainty (Fig. \ref{fig:gp_amp_vs_error}). If correlated noise amplitude can be limited to within 5-10\%, then the impact of correlated noise at R=1000 is minimal. This is not the case for R=140, which exhibits both a jump between 0\% and 10\% noise amplitude and a stronger correlation between the amplitude of correlated noise and the biosignature measurement uncertainties. 

In addition, we continue to explore the synergy between HWO and LIFE in the presence of clouds (\S \ref{sec:hwo_life}), building on previous analysis based on a cloud-free assumption~\citep{Alei2024}. We conclude that each mission will find distinct biosignatures in their wavelength coverages, e.g., O$_2$ for HWO and CO$_2$ for LIFE (see Fig. \ref{fig:model_data_hwo_life}). Together, they put tighter constraints on biosignature abundances (discussed in \S \ref{sec:life_hwo_quant} and shown in Fig. \ref{fig:corner_overlay_joint_hwo_life}). Methane (CH$_4$) detection is challenging for HWO in terms of wavelength coverage and IWA, but remains feasible for nearby habitable systems within $\sim$10 pc. For LIFE, methane detection is feasible and drives the mission requirements (\S \ref{sec:ch4}).

\noindent
{\bf{Acknowledgments}} {{This paper and a paper by Nicole Wolff and colleagues on a similar topic were submitted independently to AAS journals. Both are the result of an independent investigation into the impact of correlated noise on HWO spectral retrieval analyses.  We thank Nicole Wolff and colleagues for their collegiality.}} The authors thank the anonymous referee for insightful comments and suggestions that significantly improved the manuscript. This work is supported by the National Science Foundation under Grant No. 2143400. JW acknowledges the support through the Humboldt Research Fellowship for Experienced Researchers to carry out the research in Germany. This work has been carried out within the framework of the NCCR PlanetS supported by the Swiss National Science Foundation under grants 51NF40\_182901 and 51NF40\_205606. The authors would like to thank Romain Laugier for helpful comments and constructive conversations.

\bibliography{sample63}{}

\begin{thebibliography}{}
\expandafter\ifx\csname natexlab\endcsname\relax\def\natexlab#1{#1}\fi
\providecommand{\url}[1]{\href{#1}{#1}}
\providecommand{\dodoi}[1]{doi:~\href{http://doi.org/#1}{\nolinkurl{#1}}}
\providecommand{\doeprint}[1]{\href{http://ascl.net/#1}{\nolinkurl{http://ascl.net/#1}}}
\providecommand{\doarXiv}[1]{\href{https://arxiv.org/abs/#1}{\nolinkurl{https://arxiv.org/abs/#1}}}

\bibitem[{{Adibekyan} {et~al.}(2021){Adibekyan}, {Dorn}, {Sousa}, {Santos}, {Bitsch}, {Israelian}, {Mordasini}, {Barros}, {Delgado Mena}, {Demangeon}, {Faria}, {Figueira}, {Hakobyan}, {Oshagh}, {Soares}, {Kunitomo}, {Takeda}, {Jofr{\'e}}, {Petrucci}, \& {Martioli}}]{Adibekyan_2021}
{Adibekyan}, V., {Dorn}, C., {Sousa}, S.~G., {et~al.} 2021, Science, 374, 330, \dodoi{10.1126/science.abg8794}

\bibitem[{{Alei} {et~al.}(2024){Alei}, {Quanz}, {Konrad}, {Garvin}, {Kofman}, {Mandell}, {Angerhausen}, {Molli{\`e}re}, {Meyer}, {Robinson}, {Rugheimer}, \& {the LIFE Collaboration}}]{Alei2024}
{Alei}, E., {Quanz}, S.~P., {Konrad}, B.~S., {et~al.} 2024, \aap, 689, A245, \dodoi{10.1051/0004-6361/202450320}

\bibitem[{{Borges} {et~al.}(2024){Borges}, {Jones}, \& {Robinson}}]{Borges2024}
{Borges}, S.~R., {Jones}, G.~G., \& {Robinson}, T.~D. 2024, Astrobiology, 24, 283, \dodoi{10.1089/ast.2023.0099}

\bibitem[{{Brinkman} {et~al.}(2025){Brinkman}, {Weiss}, {Huber}, {Lee}, {Kolecki}, {Tenn}, {Zhang}, {Narayanan}, {Polanski}, {Dai}, {Bean}, {Beard}, {Brady}, {Brodheim}, {Brown}, {Chontos}, {Deich}, {Edelstein}, {Fulton}, {Giacalone}, {Gibson}, {Gilbert}, {Halverson}, {Handley}, {Hill}, {Holcomb}, {Holden}, {Householder}, {Howard}, {Isaacson}, {Kaye}, {Laher}, {Lanclos}, {Ong}, {Payne}, {Petigura}, {Pidhorodetska}, {Poppett}, {Roy}, {Rubenzahl}, {Saunders}, {Schwab}, {Seifahrt}, {Shaum}, {Sirk}, {Smith}, {Smith}, {Stef{\'a}nsson}, {St{\"u}rmer}, {Thorne}, {Turtelboom}, {Tyler}, {Valliant}, {Van Zandt}, {Walawender}, {Yee}, {Yeh}, \& {Zink}}]{Brinkman2025}
{Brinkman}, C.~L., {Weiss}, L.~M., {Huber}, D., {et~al.} 2025, \aj, 170, 109, \dodoi{10.3847/1538-3881/ade677}

\bibitem[{{Caceres} {et~al.}(2019){Caceres}, {Feigelson}, {Jogesh Babu}, {Bahamonde}, {Christen}, {Bertin}, {Meza}, \& {Cur{\'e}}}]{Caceres2019}
{Caceres}, G.~A., {Feigelson}, E.~D., {Jogesh Babu}, G., {et~al.} 2019, \aj, 158, 57, \dodoi{10.3847/1538-3881/ab26b8}

\bibitem[{{Cahoy} {et~al.}(2010){Cahoy}, {Marley}, \& {Fortney}}]{Cahoy2010}
{Cahoy}, K.~L., {Marley}, M.~S., \& {Fortney}, J.~J. 2010, \apj, 724, 189, \dodoi{10.1088/0004-637X/724/1/189}

\bibitem[{{Coker} {et~al.}(2018){Coker}, {Wang}, \& {Shaklan}}]{Coker2018}
{Coker}, C.~T., {Wang}, J., \& {Shaklan}, S. 2018, in Society of Photo-Optical Instrumentation Engineers (SPIE) Conference Series, Vol. 10698, Space Telescopes and Instrumentation 2018: Optical, Infrared, and Millimeter Wave, ed. M.~{Lystrup}, H.~A. {MacEwen}, G.~G. {Fazio}, N.~{Batalha}, N.~{Siegler}, \& E.~C. {Tong}, 106985G, \dodoi{10.1117/12.2313695}

\bibitem[{{Crouse} {et~al.}(2021){Crouse}, {Bastelberger}, {Arney}, {Domagal-Goldman}, {Virtual Planetary Laboratory (VPL)}, \& {Sellers Exoplanet Environments Collaboration (SEEC)}}]{Crouse2021}
{Crouse}, J., {Bastelberger}, S., {Arney}, G.~N., {et~al.} 2021, in Bulletin of the American Astronomical Society, Vol.~53, 0202

\bibitem[{{Currie} {et~al.}(2023){Currie}, {Stark}, {Kammerer}, {Juanola-Parramon}, \& {Meadows}}]{Currie2023}
{Currie}, M.~H., {Stark}, C.~C., {Kammerer}, J., {Juanola-Parramon}, R., \& {Meadows}, V.~S. 2023, \aj, 166, 197, \dodoi{10.3847/1538-3881/acfda7}

\bibitem[{{Ejaz} {et~al.}(2026){Ejaz}, {Dodson-Robinson}, \& {Haley}}]{Ejaz2026}
{Ejaz}, A., {Dodson-Robinson}, S., \& {Haley}, C. 2026, \aj, 171, 124, \dodoi{10.3847/1538-3881/ae2fe4}

\bibitem[{{Farr} {et~al.}(2018){Farr}, {Pope}, {Davies}, {North}, {White}, {Barrett}, {Miglio}, {Lund}, {Antoci}, {Fredslund Andersen}, {Grundahl}, \& {Huber}}]{Farr2018}
{Farr}, W.~M., {Pope}, B. J.~S., {Davies}, G.~R., {et~al.} 2018, \apjl, 865, L20, \dodoi{10.3847/2041-8213/aadfde}

\bibitem[{{Feng} {et~al.}(2018){Feng}, {Robinson}, {Fortney}, {Lupu}, {Marley}, {Lewis}, {Macintosh}, \& {Line}}]{Feng2018}
{Feng}, Y.~K., {Robinson}, T.~D., {Fortney}, J.~J., {et~al.} 2018, \aj, 155, 200, \dodoi{10.3847/1538-3881/aab95c}

\bibitem[{{Goodis Gordon} {et~al.}(2024){Goodis Gordon}, {Karalidi}, {Bott}, {Wogan}, {Arney}, {Parenteau}, {Kataria}, \& {Meadows}}]{Kenneth2024}
{Goodis Gordon}, K.~E., {Karalidi}, T., {Bott}, K.~M., {et~al.} 2024, arXiv e-prints, arXiv:2410.02194, \dodoi{10.48550/arXiv.2410.02194}

\bibitem[{{Howe} {et~al.}(2024){Howe}, {Stark}, \& {Sadleir}}]{Howe2024}
{Howe}, A.~R., {Stark}, C.~C., \& {Sadleir}, J.~E. 2024, Journal of Astronomical Telescopes, Instruments, and Systems, 10, 025008, \dodoi{10.1117/1.JATIS.10.2.025008}

\bibitem[{{Huber} {et~al.}(2025){Huber}, {Dannert}, {Laugier}, {Matsuo}, {Rutten}, {Glauser}, {Quanz}, \& {LIFE Collaboration}}]{Huber2025}
{Huber}, P.~A., {Dannert}, F.~A., {Laugier}, R., {et~al.} 2025, \aj, 170, 227, \dodoi{10.3847/1538-3881/adfb6b}

\bibitem[{{Kelkar} {et~al.}(2025){Kelkar}, {Saxena}, {Kopparapu}, \& {Monteiro}}]{Kelkar2025}
{Kelkar}, S., {Saxena}, P., {Kopparapu}, R., \& {Monteiro}, J. 2025, \psj, 6, 87, \dodoi{10.3847/PSJ/adbe7f}

\bibitem[{{Konrad} {et~al.}(2024){Konrad}, {Quanz}, {Alei}, \& {Wordsworth}}]{Konrad2024}
{Konrad}, B.~S., {Quanz}, S.~P., {Alei}, E., \& {Wordsworth}, R. 2024, \apj, 975, 13, \dodoi{10.3847/1538-4357/ad74f7}

\bibitem[{{Latouf} {et~al.}(2025){Latouf}, {Himes}, {Mandell}, {Moore}, {Kofman}, {Villanueva}, \& {Stark}}]{Latouf2025}
{Latouf}, N., {Himes}, M.~D., {Mandell}, A.~M., {et~al.} 2025, \aj, 169, 50, \dodoi{10.3847/1538-3881/ad9729}

\bibitem[{{Melton} {et~al.}(2024){Melton}, {Feigelson}, {Montalto}, {Caceres}, {Rosenswie}, \& {Abelson}}]{Melton2024}
{Melton}, E.~J., {Feigelson}, E.~D., {Montalto}, M., {et~al.} 2024, \aj, 167, 203, \dodoi{10.3847/1538-3881/ad29f1}

\bibitem[{{Mennesson} {et~al.}(2024){Mennesson}, {Belikov}, {Por}, {Serabyn}, {Ruane}, {Riggs}, {Sirbu}, {Pueyo}, {Soummer}, {Kasdin}, {Shaklan}, {Seo}, {Stark}, {Cady}, {Chen}, {Crill}, {Fogarty}, {Greenbaum}, {Guyon}, {Juanola-Parramon}, {Kern}, {Krist}, {Macintosh}, {Marx}, {Mawet}, {Prada}, {Morgan}, {Nemati}, {Pogorelyuk}, {Redmond}, {Seager}, {Siegler}, {Stapelfeldt}, {Steiger}, {Trauger}, {Wallace}, {Ygouf}, \& {Zimmerman}}]{Mennesson2024}
{Mennesson}, B., {Belikov}, R., {Por}, E., {et~al.} 2024, Journal of Astronomical Telescopes, Instruments, and Systems, 10, 035004, \dodoi{10.1117/1.JATIS.10.3.035004}

\bibitem[{{Molli{\`e}re} {et~al.}(2019){Molli{\`e}re}, {Wardenier}, {van Boekel}, {Henning}, {Molaverdikhani}, \& {Snellen}}]{Molliere2019}
{Molli{\`e}re}, P., {Wardenier}, J.~P., {van Boekel}, R., {et~al.} 2019, arXiv e-prints, arXiv:1904.11504.
\newblock \doarXiv{1904.11504}

\bibitem[{{Morgan} {et~al.}(2024){Morgan}, {Savransky}, {Turmon}, {Damiano}, {Hu}, {Mennesson}, {Mamajek}, {Robinson}, \& {Tokadjian}}]{Morgan2024}
{Morgan}, R., {Savransky}, D., {Turmon}, M., {et~al.} 2024, in Society of Photo-Optical Instrumentation Engineers (SPIE) Conference Series, Vol. 13092, Space Telescopes and Instrumentation 2024: Optical, Infrared, and Millimeter Wave, ed. L.~E. {Coyle}, S.~{Matsuura}, \& M.~D. {Perrin}, 130925M, \dodoi{10.1117/12.3020858}

\bibitem[{{Quanz} {et~al.}(2022){Quanz}, {Ottiger}, {Fontanet}, {Kammerer}, {Menti}, {Dannert}, {Gheorghe}, {Absil}, {Airapetian}, {Alei}, {Allart}, {Angerhausen}, {Blumenthal}, {Buchhave}, {Cabrera}, {Carri{\'o}n-Gonz{\'a}lez}, {Chauvin}, {Danchi}, {Dandumont}, {Defr{\'e}re}, {Dorn}, {Ehrenreich}, {Ertel}, {Fridlund}, {Garc{\'\i}a Mu{\~n}oz}, {Gasc{\'o}n}, {Girard}, {Glauser}, {Grenfell}, {Guidi}, {Hagelberg}, {Helled}, {Ireland}, {Janson}, {Kopparapu}, {Korth}, {Kozakis}, {Kraus}, {L{\'e}ger}, {Leedj{\"a}rv}, {Lichtenberg}, {Lillo-Box}, {Linz}, {Liseau}, {Loicq}, {Mahendra}, {Malbet}, {Mathew}, {Mennesson}, {Meyer}, {Mishra}, {Molaverdikhani}, {Noack}, {Oza}, {Pall{\'e}}, {Parviainen}, {Quirrenbach}, {Rauer}, {Ribas}, {Rice}, {Romagnolo}, {Rugheimer}, {Schwieterman}, {Serabyn}, {Sharma}, {Stassun}, {Szul{\'a}gyi}, {Wang}, {Wunderlich}, {Wyatt}, \& {LIFE Collaboration}}]{Quanz2022}
{Quanz}, S.~P., {Ottiger}, M., {Fontanet}, E., {et~al.} 2022, \aap, 664, A21, \dodoi{10.1051/0004-6361/202140366}

\bibitem[{{Robinson} {et~al.}(2016){Robinson}, {Stapelfeldt}, \& {Marley}}]{Robinson2016}
{Robinson}, T.~D., {Stapelfeldt}, K.~R., \& {Marley}, M.~S. 2016, \pasp, 128, 025003, \dodoi{10.1088/1538-3873/128/960/025003}

\bibitem[{{Ruffio} {et~al.}(2026){Ruffio}, {Steiger}, {Spohn}, {Macintosh}, {Mawet}, {Pueyo}, {Mennesson}, {Dacus}, {Wolff}, {Robinson}, {Hu}, {Hoch}, {Konopacky}, {Perrin}, {Savransky}, {McElwain}, {Wright}, {Wang}, \& {Chen}}]{Ruffio2026}
{Ruffio}, J.-B., {Steiger}, S., {Spohn}, C., {et~al.} 2026, arXiv e-prints, arXiv:2604.17554.
\newblock \doarXiv{2604.17554}

\bibitem[{{Schwieterman} {et~al.}(2018){Schwieterman}, {Kiang}, {Parenteau}, {Harman}, {DasSarma}, {Fisher}, {Arney}, {Hartnett}, {Reinhard}, {Olson}, {Meadows}, {Cockell}, {Walker}, {Grenfell}, {Hegde}, {Rugheimer}, {Hu}, \& {Lyons}}]{Schwieterman2018}
{Schwieterman}, E.~W., {Kiang}, N.~Y., {Parenteau}, M.~N., {et~al.} 2018, Astrobiology, 18, 663, \dodoi{10.1089/ast.2017.1729}

\bibitem[{{Speagle}(2020)}]{Speagle2020}
{Speagle}, J.~S. 2020, \mnras, 493, 3132, \dodoi{10.1093/mnras/staa278}

\bibitem[{{Steiger} {et~al.}(2024){Steiger}, {Pueyo}, {Por}, {Chen}, {Soummer}, {Pourcelot}, {Laginja}, \& {Bailey}}]{Steiger2024}
{Steiger}, S., {Pueyo}, L., {Por}, E.~H., {et~al.} 2024, in Society of Photo-Optical Instrumentation Engineers (SPIE) Conference Series, Vol. 13092, Space Telescopes and Instrumentation 2024: Optical, Infrared, and Millimeter Wave, ed. L.~E. {Coyle}, S.~{Matsuura}, \& M.~D. {Perrin}, 130921W, \dodoi{10.1117/12.3020603}

\bibitem[{{St{\"u}eken} {et~al.}(2020){St{\"u}eken}, {Som}, {Claire}, {Rugheimer}, {Scherf}, {Spro{\ss}}, {Tosi}, {Ueno}, \& {Lammer}}]{Stueken2020}
{St{\"u}eken}, E.~E., {Som}, S.~M., {Claire}, M., {et~al.} 2020, \ssr, 216, 31, \dodoi{10.1007/s11214-020-00652-3}

\bibitem[{{Wang} {et~al.}(2018){Wang}, {Mawet}, {Hu}, {Ruane}, {Delorme}, \& {Klimovich}}]{Wang2018JATIS}
{Wang}, J., {Mawet}, D., {Hu}, R., {et~al.} 2018, Journal of Astronomical Telescopes, Instruments, and Systems, 4, 035001, \dodoi{10.1117/1.JATIS.4.3.035001}

\bibitem[{{Wang} {et~al.}(2017){Wang}, {Mawet}, {Ruane}, {Hu}, \& {Benneke}}]{Wang2017a}
{Wang}, J., {Mawet}, D., {Ruane}, G., {Hu}, R., \& {Benneke}, B. 2017, \aj, 153, 183, \dodoi{10.3847/1538-3881/aa6474}

\end{thebibliography}
\bibliographystyle{aasjournal}



\begin{deluxetable}{llll}
\tablewidth{0pt}
\centering
\tablecaption{Parameters Used in the Bayesian Inference \label{tab:retrieval_params}}
\tablehead{
\colhead{\textbf{Parameter}} &
\colhead{\textbf{Description}} &
\colhead{\textbf{Input}} &
\colhead{\textbf{Prior}}
}

\startdata
$\log P_0$ (bar) & Surface pressure & $\log(1)$ & $[-2, 2]$ \\
$\log \mathrm{H_2O}$ & Water vapor mixing ratio & $\log(3 \times 10^{-3})$ & $[-8, -1]$ \\
$\log \mathrm{CO_2}$$^\ast$ & Carbon dioxide mixing ratio & $\log(4 \times 10^{-4})$ & $[-8, -1]$ \\
$\log \mathrm{O_2}$ & Molecular oxygen mixing ratio & $\log(0.21)$ & $[-8, 0]$ \\
$\log \mathrm{O_3}$$^\ast$ & Ozone mixing ratio & $\log(7 \times 10^{-7})$ & $[-12, -3]$ \\
$\log R_p$ ($R_\oplus$) & Planet radius & $\log(1)$ & $[-1, 1]$ \\
$\log g$ (g$_\oplus$) & Scaled surface gravity & $\log(1)$ & $[-1, 1]$ \\
$\log A_s$ & Albedo & $\log(0.3)$ & $[-2, 0]$ \\
$\log f_c$$^\ast$ & Cloud fraction & $\log(0.5)$ & $[-3, 0]$ \\
$\log \kappa_{350\,\mathrm{nm}}$$^\ast$ & Cloud opacity at 350 nm & $\log(0.01)$ & $[-4, 0]$ \\
$p_{\rm cl}$$^\ast$ & Cloud opacity power-law coefficient & $-3.0$ & $[-5, 0]$ \\
$\alpha$$^\ast$ & Temperature--pressure profile slope & $0.2$ & $[0, 1]$ \\
$\log \lambda_{\rm GP}$ ($\mu$m) & Correlation length & $-1.5$ & $[-5, 0]$ \\
$\log \sigma_{\rm GP}$  & Correlated amplitude & $-0.5$ or $-1.0$ & $[-5, 0]$ \\
\enddata
\tablecomments{{{Parameters without $\ast$ are used in HWO-only (\S \ref{sec:major_findings}) and HWO-LIFE analyses (\S \ref{sec:hwo_life}).}} Parameters marked with $\ast$ are only used in HWO-LIFE analyses (\S \ref{sec:hwo_life}).  }
\end{deluxetable}


\begin{deluxetable}{lccccccccccc}
\rotate
\tabletypesize{\tiny}
\tablewidth{0pt}
\tablecaption{Posterior constraints for the R=140 and non-GP retrieval set.\label{tab:r140_snr20_nogp}}
\tablehead{
\colhead{Parameter} & \colhead{Input} & \colhead{0} & \colhead{1} & \colhead{2} & \colhead{3} & \colhead{4} & \colhead{5} & \colhead{6} & \colhead{7} & \colhead{8} & \colhead{9} \\
}
\startdata
$\log P_0$ (bar) & $\log(1)$ & \textcolor{orange}{$0.41^{+0.41}_{-0.37}$} & $-0.07^{+0.41}_{-0.33}$ & $-0.03^{+0.37}_{-0.32}$ & $0.39^{+0.44}_{-0.42}$ & $0.36^{+0.46}_{-0.39}$ & $-0.15^{+0.36}_{-0.34}$ & $0.21^{+0.35}_{-0.31}$ & $-0.04^{+0.42}_{-0.38}$ & $-0.17^{+0.31}_{-0.28}$ & \textcolor{orange}{$0.43^{+0.33}_{-0.34}$} \\
$\log \mathrm{H_2O}$ & $\log(3 \times 10^{-3})$ & \textcolor{orange}{$-3.19^{+0.51}_{-0.59}$} & $-2.47^{+0.39}_{-0.52}$ & $-2.52^{+0.49}_{-0.62}$ & $-3.07^{+0.56}_{-0.71}$ & $-2.67^{+0.53}_{-0.83}$ & $-2.16^{+0.59}_{-0.68}$ & $-2.83^{+0.46}_{-0.62}$ & $-2.58^{+0.53}_{-0.66}$ & $-2.55^{+0.43}_{-0.47}$ & $-2.60^{+0.36}_{-0.47}$ \\
$\log \mathrm{O_2}$ & $\log(0.21)$ & \textcolor{orange}{$-1.42^{+0.50}_{-0.63}$} & $-0.59^{+0.33}_{-0.52}$ & $-0.75^{+0.43}_{-0.66}$ & $-1.26^{+0.55}_{-0.76}$ & $-0.92^{+0.51}_{-0.95}$ & $-1.16^{+0.66}_{-1.26}$ & $-0.92^{+0.39}_{-0.63}$ & $-0.87^{+0.50}_{-0.73}$ & $-0.66^{+0.32}_{-0.45}$ & $-0.70^{+0.30}_{-0.51}$ \\
$R_p$ ($R_{\oplus}$) & $\log(1)$ & $-0.09^{+0.12}_{-0.08}$ & $-0.04^{+0.14}_{-0.11}$ & $-0.07^{+0.17}_{-0.11}$ & $-0.07^{+0.14}_{-0.10}$ & $-0.05^{+0.17}_{-0.12}$ & $0.03^{+0.20}_{-0.15}$ & \textcolor{orange}{$-0.14^{+0.12}_{-0.07}$} & $0.03^{+0.18}_{-0.13}$ & \textcolor{orange}{$-0.14^{+0.10}_{-0.07}$} & $-0.06^{+0.14}_{-0.12}$ \\
$\log g$ (cm s$^{-2}$) & $\log(1)$ & $-0.16^{+0.66}_{-0.56}$ & $-0.17^{+0.67}_{-0.57}$ & $-0.03^{+0.63}_{-0.61}$ & $0.05^{+0.61}_{-0.65}$ & $0.41^{+0.42}_{-0.68}$ & $0.10^{+0.58}_{-0.67}$ & $-0.14^{+0.65}_{-0.58}$ & $-0.08^{+0.67}_{-0.61}$ & $-0.47^{+0.60}_{-0.38}$ & $0.51^{+0.35}_{-0.59}$ \\
$\log A_s$ & $\log(0.3)$ & $-0.37^{+0.17}_{-0.25}$ & $-0.45^{+0.22}_{-0.29}$ & $-0.37^{+0.22}_{-0.34}$ & $-0.40^{+0.20}_{-0.29}$ & $-0.41^{+0.25}_{-0.34}$ & $-0.58^{+0.31}_{-0.42}$ & \textcolor{orange}{$-0.23^{+0.15}_{-0.25}$} & $-0.60^{+0.26}_{-0.36}$ & \textcolor{orange}{$-0.23^{+0.15}_{-0.21}$} & $-0.39^{+0.25}_{-0.28}$ \\
\enddata
\tablecomments{{\bf{Retrieved values with 1-$\sigma$ upper and lower limits.}} The numbers in the header refer to the number of the mock data being analyzed. Orange entries are more than $1\sigma$ from the default value, red entries are more than $2\sigma$ from the default value.}
\end{deluxetable}


\begin{deluxetable}{lccccccccccc}
\rotate
\tabletypesize{\tiny}
\tablewidth{0pt}
\tablecaption{Posterior constraints for the R=140 and GP retrieval set.\label{tab:r140_snr20}}
\tablehead{
\colhead{Parameter} & \colhead{Input} & \colhead{0} & \colhead{1} & \colhead{2} & \colhead{3} & \colhead{4} & \colhead{5} & \colhead{6} & \colhead{7} & \colhead{8} & \colhead{9} \\
}
\startdata
$\log P_0$ (bar) & $\log(1)$ & \textcolor{orange}{$0.85^{+0.68}_{-0.62}$} & \textcolor{orange}{$0.77^{+0.64}_{-0.63}$} & $0.67^{+0.73}_{-0.69}$ & \textcolor{orange}{$1.43^{+0.40}_{-1.18}$} & \textcolor{orange}{$1.18^{+0.51}_{-0.60}$} & $0.24^{+0.75}_{-0.57}$ & $-0.03^{+0.80}_{-0.56}$ & $0.28^{+0.95}_{-0.53}$ &  & \textcolor{orange}{$0.93^{+0.72}_{-0.86}$} \\
$\log \mathrm{H_2O}$ & $\log(3 \times 10^{-3})$ & $-3.37^{+1.00}_{-0.90}$ & $-3.39^{+0.83}_{-0.98}$ & \textcolor{orange}{$-3.58^{+0.98}_{-1.05}$} & \textcolor{orange}{$-4.02^{+1.34}_{-1.09}$} & \textcolor{red}{$-4.01^{+0.72}_{-0.62}$} & $-2.77^{+0.96}_{-1.20}$ & $-2.53^{+0.93}_{-1.27}$ & $-2.72^{+0.81}_{-1.37}$ &  & $-3.72^{+1.26}_{-0.94}$ \\
$\log \mathrm{O_2}$ & $\log(0.21)$ & \textcolor{orange}{$-2.34^{+1.13}_{-1.10}$} & $-1.56^{+0.85}_{-1.09}$ & $-1.75^{+0.92}_{-1.09}$ & \textcolor{orange}{$-1.91^{+0.97}_{-1.10}$} & \textcolor{orange}{$-1.78^{+0.66}_{-0.61}$} & $-1.58^{+0.95}_{-1.38}$ & $-1.99^{+1.26}_{-3.98}$ & $-1.08^{+0.74}_{-1.51}$ &  & $-1.81^{+1.18}_{-0.99}$ \\
$R_p$ ($R_{\oplus}$) & $\log(1)$ & $-0.13^{+0.19}_{-0.09}$ & $-0.12^{+0.21}_{-0.11}$ & $-0.09^{+0.20}_{-0.10}$ & $-0.24^{+0.32}_{-0.05}$ & \textcolor{orange}{$-0.19^{+0.14}_{-0.08}$} & $-0.05^{+0.23}_{-0.13}$ & $-0.05^{+0.27}_{-0.15}$ & $-0.00^{+0.28}_{-0.18}$ &  & \textcolor{orange}{$-0.18^{+0.17}_{-0.07}$} \\
$\log g$ (cm s$^{-2}$) & $\log(1)$ & $0.28^{+0.52}_{-0.71}$ & $0.00^{+0.67}_{-0.67}$ & $-0.12^{+0.72}_{-0.60}$ & \textcolor{orange}{$-0.72^{+0.53}_{-0.20}$} & $-0.06^{+0.65}_{-0.58}$ & $0.08^{+0.62}_{-0.68}$ & $-0.02^{+0.68}_{-0.66}$ & $-0.13^{+0.71}_{-0.60}$ &  & $-0.29^{+0.78}_{-0.48}$ \\
$\log A_s$ & $\log(0.3)$ & $-0.38^{+0.26}_{-0.46}$ & $-0.52^{+0.34}_{-0.54}$ & $-0.48^{+0.33}_{-0.54}$ & $-1.11^{+0.65}_{-0.59}$ & $-0.50^{+0.30}_{-0.49}$ & $-0.45^{+0.31}_{-0.49}$ & $-0.50^{+0.35}_{-0.57}$ & $-0.77^{+0.47}_{-0.61}$ &  & $-0.34^{+0.23}_{-0.47}$ \\
$\log \lambda_{\rm{GP}}$ ($\mu$m) & $-1.5$ & $-1.53^{+0.07}_{-0.08}$ & \textcolor{orange}{$-1.36^{+0.06}_{-0.08}$} & $-1.50^{+0.05}_{-0.06}$ & $-1.55^{+0.07}_{-0.08}$ & $-1.42^{+0.08}_{-0.08}$ & $-1.55^{+0.08}_{-0.09}$ & $-1.43^{+0.08}_{-0.08}$ & \textcolor{orange}{$-1.36^{+0.07}_{-0.09}$} &  & \textcolor{orange}{$-1.42^{+0.07}_{-0.08}$} \\
$\log \sigma_{\rm{GP}}$ & $-0.5$ & $-0.52^{+0.13}_{-0.11}$ & \textcolor{red}{$-0.20^{+0.11}_{-0.12}$} & $-0.49^{+0.13}_{-0.11}$ & \textcolor{orange}{$-0.36^{+0.14}_{-0.12}$} & $-0.38^{+0.15}_{-0.13}$ & \textcolor{orange}{$-0.70^{+0.15}_{-0.13}$} & \textcolor{orange}{$-0.33^{+0.13}_{-0.12}$} & $-0.36^{+0.15}_{-0.13}$ &  & \textcolor{orange}{$-0.33^{+0.14}_{-0.12}$} \\
\enddata
\tablecomments{Retrieved values with 1-$\sigma$ upper and lower limits. The numbers in the header refer to the number of the mock data being analyzed. Run \#8 was not converged likely due to the large GP amplitude and/or the low spectral resolution. Orange entries are more than $1\sigma$ from the default value, red entries are more than $2\sigma$ from the default value. Missing files are left blank.}
\end{deluxetable}


\begin{deluxetable}{lccccccccccc}
\rotate
\tabletypesize{\tiny}
\tablewidth{0pt}
\tablecaption{Posterior constraints for the R=1000 and non-GP retrieval set.\label{tab:r1000_snr7p5_nogp}}
\tablehead{
\colhead{Parameter} & \colhead{Input} & \colhead{0} & \colhead{1} & \colhead{2} & \colhead{3} & \colhead{4} & \colhead{5} & \colhead{6} & \colhead{7} & \colhead{8} & \colhead{9} \\
}
\startdata
$\log P_0$ (bar) & $\log(1)$ & \textcolor{orange}{$0.27^{+0.26}_{-0.25}$} & $-0.04^{+0.28}_{-0.26}$ & $-0.09^{+0.29}_{-0.27}$ & $0.16^{+0.25}_{-0.26}$ & \textcolor{orange}{$0.35^{+0.34}_{-0.35}$} & \textcolor{orange}{$0.44^{+0.33}_{-0.29}$} & $0.19^{+0.35}_{-0.31}$ & \textcolor{orange}{$-0.34^{+0.25}_{-0.20}$} & $0.21^{+0.27}_{-0.27}$ & $0.21^{+0.25}_{-0.25}$ \\
$\log \mathrm{H_2O}$ & $\log(3 \times 10^{-3})$ & $-2.60^{+0.35}_{-0.51}$ & $-2.59^{+0.34}_{-0.41}$ & $-2.38^{+0.30}_{-0.39}$ & $-2.42^{+0.33}_{-0.52}$ & $-2.73^{+0.45}_{-0.70}$ & $-2.56^{+0.35}_{-0.67}$ & $-2.54^{+0.44}_{-0.66}$ & \textcolor{orange}{$-2.21^{+0.24}_{-0.28}$} & $-2.78^{+0.45}_{-0.56}$ & $-2.40^{+0.35}_{-0.49}$ \\
$\log \mathrm{O_2}$ & $\log(0.21)$ & $-0.78^{+0.33}_{-0.53}$ & $-0.75^{+0.30}_{-0.42}$ & $-0.45^{+0.24}_{-0.38}$ & $-0.56^{+0.28}_{-0.51}$ & $-0.95^{+0.47}_{-0.76}$ & $-0.78^{+0.38}_{-0.76}$ & $-0.83^{+0.45}_{-0.74}$ & \textcolor{orange}{$-0.35^{+0.19}_{-0.25}$} & $-1.12^{+0.44}_{-0.58}$ & $-0.71^{+0.35}_{-0.51}$ \\
$R_p$ ($R_{\oplus}$) & $\log(1)$ & $-0.13^{+0.13}_{-0.08}$ & \textcolor{red}{$-0.18^{+0.08}_{-0.05}$} & $0.03^{+0.14}_{-0.12}$ & $-0.03^{+0.15}_{-0.13}$ & $-0.09^{+0.15}_{-0.10}$ & $0.01^{+0.14}_{-0.15}$ & $-0.03^{+0.16}_{-0.13}$ & $0.01^{+0.11}_{-0.08}$ & $-0.02^{+0.15}_{-0.11}$ & $-0.05^{+0.15}_{-0.12}$ \\
$\log g$ (cm s$^{-2}$) & $\log(1)$ & $0.33^{+0.43}_{-0.59}$ & $-0.25^{+0.60}_{-0.48}$ & $-0.04^{+0.59}_{-0.57}$ & $0.32^{+0.45}_{-0.67}$ & $0.32^{+0.47}_{-0.69}$ & \textcolor{orange}{$0.59^{+0.30}_{-0.57}$} & $0.29^{+0.49}_{-0.67}$ & $-0.48^{+0.52}_{-0.36}$ & $0.03^{+0.57}_{-0.61}$ & $0.38^{+0.41}_{-0.59}$ \\
$\log A_s$ & $\log(0.3)$ & \textcolor{orange}{$-0.25^{+0.17}_{-0.27}$} & \textcolor{red}{$-0.14^{+0.10}_{-0.16}$} & $-0.57^{+0.24}_{-0.28}$ & $-0.45^{+0.28}_{-0.31}$ & $-0.34^{+0.21}_{-0.30}$ & $-0.54^{+0.30}_{-0.29}$ & $-0.45^{+0.26}_{-0.33}$ & $-0.55^{+0.17}_{-0.22}$ & $-0.49^{+0.22}_{-0.30}$ & $-0.40^{+0.25}_{-0.31}$ \\
\enddata
\tablecomments{Retrieved values with 1-$\sigma$ upper and lower limits. The numbers in the header refer to the number of the mock data being analyzed. Orange entries are more than $1\sigma$ from the default value, red entries are more than $2\sigma$ from the default value. }
\end{deluxetable}


\begin{deluxetable}{lccccccccccc}
\rotate
\tabletypesize{\tiny}
\tablewidth{0pt}
\tablecaption{Posterior constraints for the R=1000 and GP retrieval set.\label{tab:r1000_snr7p5_gp}}
\tablehead{
\colhead{Parameter} & \colhead{Input} & \colhead{1} & \colhead{2} & \colhead{3} & \colhead{4} & \colhead{5} & \colhead{6} & \colhead{7} & \colhead{8} & \colhead{9} & \colhead{10} \\
}
\startdata
$\log P_0$ (bar) & $\log(1)$ & \textcolor{orange}{$0.46^{+0.39}_{-0.34}$} & $0.02^{+0.31}_{-0.31}$ & $0.10^{+0.39}_{-0.35}$ & \textcolor{orange}{$0.73^{+0.42}_{-0.47}$} & $0.35^{+0.38}_{-0.32}$ & $0.31^{+0.50}_{-0.41}$ & $0.19^{+0.36}_{-0.34}$ & $0.25^{+0.42}_{-0.34}$ & $-0.15^{+0.36}_{-0.34}$ & \textcolor{orange}{$0.59^{+0.52}_{-0.55}$} \\
$\log \mathrm{H_2O}$ & $\log(3 \times 10^{-3})$ & $-2.91^{+0.54}_{-0.70}$ & $-2.65^{+0.51}_{-0.61}$ & $-2.51^{+0.49}_{-0.75}$ & \textcolor{orange}{$-3.42^{+0.73}_{-0.77}$} & $-3.20^{+0.64}_{-0.79}$ & $-2.94^{+0.58}_{-0.78}$ & $-2.44^{+0.57}_{-0.68}$ & $-2.62^{+0.47}_{-0.83}$ & $-2.39^{+0.47}_{-0.58}$ & $-3.22^{+0.67}_{-0.75}$ \\
$\log \mathrm{O_2}$ & $\log(0.21)$ & $-1.14^{+0.56}_{-0.76}$ & $-0.87^{+0.47}_{-0.63}$ & $-0.70^{+0.44}_{-0.77}$ & \textcolor{orange}{$-1.80^{+0.79}_{-0.82}$} & $-1.25^{+0.61}_{-0.82}$ & $-1.17^{+0.55}_{-0.82}$ & $-1.01^{+0.55}_{-0.70}$ & $-0.70^{+0.48}_{-0.96}$ & $-0.79^{+0.43}_{-0.58}$ & $-1.50^{+0.76}_{-0.86}$ \\
$R_p$ ($R_{\oplus}$) & $\log(1)$ & $-0.05^{+0.21}_{-0.14}$ & $-0.05^{+0.20}_{-0.13}$ & $-0.07^{+0.22}_{-0.13}$ & $-0.09^{+0.18}_{-0.10}$ & $-0.12^{+0.16}_{-0.09}$ & $-0.01^{+0.20}_{-0.16}$ & $-0.04^{+0.24}_{-0.16}$ & $0.00^{+0.23}_{-0.15}$ & $0.00^{+0.22}_{-0.18}$ & $-0.10^{+0.17}_{-0.10}$ \\
$\log g$ (cm s$^{-2}$) & $\log(1)$ & $0.30^{+0.48}_{-0.65}$ & $-0.16^{+0.64}_{-0.56}$ & $-0.03^{+0.66}_{-0.64}$ & $0.17^{+0.58}_{-0.71}$ & $-0.25^{+0.71}_{-0.53}$ & $-0.28^{+0.67}_{-0.49}$ & $0.03^{+0.61}_{-0.64}$ & $0.47^{+0.39}_{-0.70}$ & $-0.41^{+0.61}_{-0.42}$ & $0.25^{+0.52}_{-0.71}$ \\
$\log A_s$ & $\log(0.3)$ & $-0.48^{+0.34}_{-0.46}$ & $-0.43^{+0.30}_{-0.44}$ & $-0.42^{+0.30}_{-0.46}$ & $-0.34^{+0.24}_{-0.41}$ & $-0.31^{+0.22}_{-0.37}$ & $-0.71^{+0.43}_{-0.47}$ & $-0.52^{+0.36}_{-0.54}$ & $-0.47^{+0.33}_{-0.48}$ & $-0.59^{+0.41}_{-0.52}$ & $-0.31^{+0.22}_{-0.36}$ \\
$\log \lambda_{\rm{GP}}$ ($\mu$m) & $-1.5$ & \textcolor{orange}{$-1.41^{+0.08}_{-0.09}$} & $-1.54^{+0.06}_{-0.07}$ & \textcolor{orange}{$-1.43^{+0.05}_{-0.05}$} & $-1.42^{+0.08}_{-0.09}$ & $-1.52^{+0.08}_{-0.08}$ & $-1.54^{+0.06}_{-0.07}$ & $-1.51^{+0.06}_{-0.06}$ & \textcolor{orange}{$-1.41^{+0.05}_{-0.06}$} & $-1.49^{+0.06}_{-0.07}$ & \textcolor{orange}{$-1.40^{+0.06}_{-0.07}$} \\
$\log \sigma_{\rm{GP}}$ & $-0.5$ & $-0.45^{+0.17}_{-0.13}$ & $-0.55^{+0.13}_{-0.11}$ & $-0.41^{+0.12}_{-0.11}$ & $-0.38^{+0.16}_{-0.13}$ & \textcolor{orange}{$-0.66^{+0.15}_{-0.13}$} & \textcolor{orange}{$-0.33^{+0.12}_{-0.11}$} & \textcolor{red}{$-0.23^{+0.12}_{-0.11}$} & \textcolor{orange}{$-0.35^{+0.13}_{-0.11}$} & $-0.38^{+0.13}_{-0.11}$ & \textcolor{orange}{$-0.27^{+0.13}_{-0.13}$} \\
\enddata
\tablecomments{Retrieved values with 1-$\sigma$ upper and lower limits. The numbers in the header refer to the number of the mock data being analyzed. Orange entries are more than $1\sigma$ from the default value, red entries are more than $2\sigma$ from the default value. Missing files are left blank.}
\end{deluxetable}


\begin{deluxetable*}{llccc}
\tabletypesize{\scriptsize}
\tablewidth{0pt}
\tablecaption{Posterior constraints for the HWO, LIFE, and joint HWO+LIFE retrievals.\label{tab:hwo_life_joint_constraints}}
\tablehead{
\colhead{Parameter} &
\colhead{Input} &
\colhead{HWO} &
\colhead{LIFE} &
\colhead{HWO+LIFE} \\
}
\startdata
$\log P_0$ (bar) & $\log(1)$ & $-0.07^{+0.54}_{-0.71}$ & $-0.14^{+0.20}_{-0.11}$ & \textcolor{orange}{$-0.19^{+0.14}_{-0.11}$} \\
$\log \mathrm{H_2O}$ & $\log(3 \times 10^{-3})$ & \textcolor{orange}{$-4.19^{+1.48}_{-2.30}$} & \textcolor{orange}{$-1.89^{+0.56}_{-0.60}$} & $-2.78^{+0.43}_{-0.42}$ \\
$\log \mathrm{CO_2}$ & $\log(4 \times 10^{-4})$ & $-4.45^{+2.29}_{-2.34}$ & $-3.63^{+0.63}_{-0.66}$ & \textcolor{orange}{$-3.99^{+0.50}_{-0.49}$} \\
$\log \mathrm{O_2}$ & $\log(0.21)$ & $-0.75^{+0.46}_{-1.09}$ & \textcolor{orange}{$-3.86^{+2.33}_{-2.55}$} & $-0.78^{+0.31}_{-0.89}$ \\
$\log \mathrm{O_3}$ & $\log(7 \times 10^{-7})$ & $-7.35^{+2.98}_{-3.19}$ & $-6.27^{+0.54}_{-0.54}$ & \textcolor{orange}{$-6.77^{+0.38}_{-0.27}$} \\
$\log R_p$ ($R_{\oplus}$) & $\log(1)$ & \textcolor{orange}{$0.20^{+0.12}_{-0.16}$} & $0.03^{+0.04}_{-0.05}$ & $0.04^{+0.05}_{-0.05}$ \\
$\log g$ ($g_{\oplus}$) & $\log(1)$ & $0.10^{+0.60}_{-0.72}$ & $-0.22^{+0.51}_{-0.48}$ & \textcolor{red}{$-0.82^{+0.30}_{-0.14}$} \\
$\log f_c$ & $\log(0.5)$ & \textcolor{orange}{$-1.29^{+0.65}_{-1.13}$} & $-1.05^{+0.84}_{-1.30}$ & \textcolor{orange}{$-0.86^{+0.40}_{-0.85}$} \\
$\log \kappa_{350\,\mathrm{nm}}$ & $\log(0.01)$ & $-2.69^{+1.06}_{-0.90}$ & $-1.69^{+1.19}_{-1.45}$ & \textcolor{orange}{$-2.80^{+0.56}_{-0.66}$} \\
$p_{\rm cl}$ & $-3.0$ & $-2.28^{+1.62}_{-1.79}$ & $-2.86^{+1.49}_{-1.37}$ & $-1.68^{+1.03}_{-1.57}$ \\
$\alpha$ & $0.2$ & $0.44^{+0.35}_{-0.30}$ & \textcolor{orange}{$0.29^{+0.11}_{-0.06}$} & \textcolor{red}{$0.29^{+0.09}_{-0.05}$} \\
$\log \lambda_{\rm GP}$ ($\mu$m) & $-1.5$ & $-1.49^{+0.04}_{-0.04}$ & \nodata & $-1.50^{+0.04}_{-0.04}$ \\
$\log \sigma_{\rm GP}$ & $-0.5$ & $-0.48^{+0.10}_{-0.08}$ & \nodata & $-0.46^{+0.11}_{-0.10}$ \\
\enddata
\tablecomments{Retrieved values with 1-$\sigma$ upper and lower limits. Orange entries have input values outside the central 68\% credible interval, and red entries have input values outside the central 95\% credible interval. Parameters not included in a retrieval are marked with \texttt{\nodata}.}
\end{deluxetable*}

\newpage

\clearpage

\end{CJK*}
 
\end{document}